\documentclass[epj]{svjour}
\usepackage{graphics}
\usepackage{graphicx}
\begin{document}
\title{Polarization in Hyperon Photo- and Electro- Production}
\author{Reinhard Schumacher}
%
\mail{schumacher@cmu.edu}          
\institute{Department of Physics, Carnegie Mellon University, Pittsburgh, PA 15213, USA}
\date{Invited paper at ``NStar 2007'', 10-22-07, Accepted for EPJ A, 1-21-08 }
%
\abstract{ Multiple polarization observables must be measured to
access the amplitude structure of pseudoscalar meson photoproduction
off the proton.  The hyperon-producing reactions are especially
attractive to study, since the weak decays allow straightforward
measurement of the induced and recoil polarization observables.  In
this paper we emphasize $\gamma + p \rightarrow K^+ + \Lambda$,
discussing recent measurements of $C_x$, $C_z$, and $P$ for this
reaction. An empirical constraint on the helicity amplitudes is
obtained.  A simplified model involving spin-flip and spin non-flip
amplitudes is presented.  Finally, a semi-classical model of how the
polarization may arise is presented.
} 
\maketitle
\section{Introduction}
\label{intro}

There are exactly four independent amplitudes that contribute to
pseudoscalar meson photoproduction on the nucleon, and they represent
the totality of what may be gleaned from a set of measurements of any
given reaction channel.  All possible observable quantities are
encoded by bilinear combinations of the four amplitudes, leading to
sixteen observables quantities.  A complete, model-inde\-pendent
description of a reaction is achieved if enough experimental
information is on hand to uniquely determine these amplitudes, meaning
that no assumptions are made about the underlying dynamics of the
process.

The amplitudes may be picked in a variety of ways, but the
commonly-adopted formulation uses helicity amplitudes, which
correspond to basis states in which all the particles in the reaction
have well-defined spin-projections along their direction of motion,
and the reaction has well-defined total angular momentum states
$J$~\cite{jacobwick}.  At very high energies, or in the limit of
massless particles, helicity of particles is conserved.  In this
discussion we use the helicity amplitudes in the notation of Barker,
Donnachie, and Storrow (BDS)~\cite{bds}.  These will be itemized
below.

\begin{figure}
\resizebox{0.50\textwidth}{!}{\includegraphics{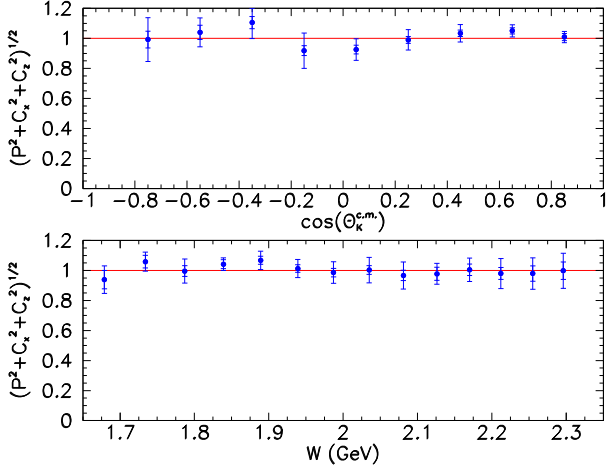}}
\caption{ 
(color online) 
Magnitude of the polarization of the
$\Lambda$ hyperon when averaging over all measured energies and given
as a function of angle (top), and alternatively when averaged over all
angles given as a function of c.m. energy (bottom).  The magnitude is
consistent with unity everywhere. The vector components $C_x$, $C_z$,
and $P$ are discussed in the text and given in terms of helicity
amplitudes in Table~\ref{obstable}.  
}
\label{fig:lambda}       
\end{figure}

Recently \cite{bradford}, the CLAS collaboration published results for
the polarization transfer from circularly polarized photons to the
recoiling hyperon in $\gamma+ p\to K^+ + \Lambda$ on an unpolarized
proton target.  The energies ranged from threshold at $W=1.6$ GeV up
to about $W=2.4$ GeV.  Two observations were made that indicate that
this reaction proceeds very far from the helicity conserving limit.
First, the spin polarization of the photons was, to first order,
transferred to the hyperons along the same axis as the photon
polarization, which we will refer to as the ``$z$-spin axis''.
Second, combining these results with earlier results published for the
induced polarization~\cite{mcnabb}, the total magnitude of the hyperon
polarization vector was unity, irrespective of the energy or angle of
the production~\cite{schumachercxcz}.  This latter result is shown in
Fig.~\ref{fig:lambda}, while Ref.~\cite{bradford} exhibits the recoil
polarization observables components in detail.  Even at energies and
angles where the photon polarization was not transferred fully along
the $z$-spin axis, the magnitude was preserved while the direction of
the polarization vector was changed.

In this paper we will discuss some implications of these observations
for other observables that have been measured already at
GRAAL~\cite{lleres}, and LEPS~\cite{zegers}, or which are due to be
measured at CLAS~\cite{frost} or ELSA~\cite{bonn}.  We begin with a
brief review of the formalism that defined the observables in a
helicity basis.  We will also discuss a toy model that arises under
some simplifying assumptions.  After discussing the measurements, we
then show what conclusions may be drawn about the amplitude structure
of this reaction.  Finally we discuss a heuristic semi-classical
picture of how hyperon polarization could arise in this reaction.

\section{Formalism}
\label{sec:1}

\subsection{Helicity basis observables}
\label{sec:helicity}

\begin{figure}
\includegraphics[scale=0.35,angle=-90.0]{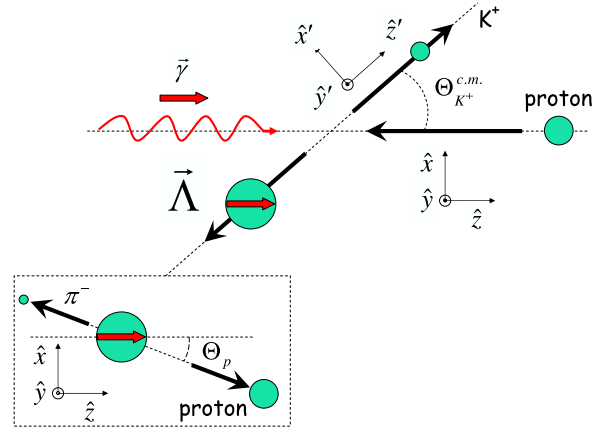}
\caption{ 
(color online) In the overall reaction center of mass frame, the
coordinate system can be oriented along the outgoing $K^+$ meson
$\{\hat{x}^\prime,\hat{y}^\prime,\hat{z}^\prime\}$ (helicity basis),
or along the incident photon direction $\{\hat{x},\hat{y},\hat{z}\}$
($z$-spin basis).  The dotted box represents the rest frame of the
hyperon, and the coordinate system used for specifying the
polarization components. The short thick red arrows represent polarization
vectors.
}
\label{fig:axes}       
\end{figure}

Considerable progress has been made to working out how many
measurements of related observables are needed to uniquely define the
four relevant amplitudes
~\cite{bds,goldstein,worden,tabakin,Artru:2006xf}.  The intricate
algebraic relations among the observables and amplitudes made this a
challenging exercise.  Table \ref{obstable} lists the set of 16
observables relevant to this discussion.  It is now known that all
three of the single spin observables and the cross section must be
measured, plus a judiciously-chosen set of 4 double polarization
observables~\cite{tabakin}.  The observables are given in terms of the
helicity basis, in which quantization axes are taken along the
momentum direction of initial and final state particles, shown as the
``primed'' variables in Fig.~\ref{fig:axes}.

In the helicity basis, the meaning of the amplitudes is as
follows. Consider $f_{0,\Lambda;\gamma,N}$ to be the amplitude that
takes a photon in helicity state $\gamma$ and proton in helicity state
$N$ into a final state with a spinless meson (kaon) and a recoiling
baryon ($\Lambda$ hyperon) in helicity state $\Lambda$.  (The notation
of Goldstein {\it et al.}~\cite{goldstein} has been adopted.)  There
are eight such combinations, forming a $2 \times 4$ matrix connecting
the $2\times2$ direct product space of helicities in the initial state
to the 2-fold final helicity space.  Parity considerations reduce the
number to four, and the following naming scheme is defined:
\begin{eqnarray}
f_{0,-\frac{1}{2};1,\frac{1}{2}} &= N &= -f_{0,\frac{1}{2};-1,-\frac{1}{2}},   \\
f_{0,\frac{1}{2};1,-\frac{1}{2}} &= D &= -f_{0,-\frac{1}{2};-1,\frac{1}{2}}
\label{eq:f1} \\
f_{0,-\frac{1}{2};1,-\frac{1}{2}} &= S_1 &=
+f_{0,\frac{1}{2};-1,\frac{1}{2}},   \\
f_{0,\frac{1}{2};1,\frac{1}{2}} &= S_2 &= +f_{0,-\frac{1}{2};-1,-\frac{1}{2}}
\label{eq:f2} 
\end{eqnarray}
The overall helicity flip of each of the amplitudes (defined as
$|\gamma - (N - \Lambda)|$) is either none ($N$), single ($S_1$ and
$S_2$) or double ($D$).  It is straightforward to compute the
observables in terms of these amplitudes, and the results are as
summarized in Table~\ref{obstable}.  Below, we present what the CLAS
measurements imply for a constraint among these amplitudes.

\begin{table*}[htpb]
\begin{center}
\begin{tabular}{|c|c|c|c|c|}
\hline
Observable & Helicity Representation & Beam & Target & Hyperon\\
\hline
\hline
\multicolumn{4}{c}{Single Polarization \& Cross Section}\\
\hline
$|A|^2 \sim d\sigma /dt$& $|N|^2 + |S|^2 + |S|^2 + |D|^2$   & -        & -                  & -\\
$\Sigma \frac{d\sigma}{dt}$ & $2 Re (S_1^*S_2 - N D^*) $    & linear   & -                  & -\\
$T \frac{d\sigma}{dt}$      &$2 Im (S_1^*N - S_2 D^*) $     & -        & transverse         & -\\
$P \frac{d\sigma}{dt}$      & $2 Im(S_2N^* - S_1 D^*)$      & -        & -                  & along $y^\prime$\\
\hline
\hline
\multicolumn{4}{c}{Beam and Target Polarization}\\
\hline
$G \frac{d\sigma}{dt}$ &$-2Im(S_1S_2^* + N D^*)$            & linear   & along $z^\prime $  & -\\
$H \frac{d\sigma}{dt}$ &$-2Im(S_2N^* + S_1D^*)$             & linear   & along $x^\prime $  & -\\
$E \frac{d\sigma}{dt}$ &$|S_1|^2 - |S_2|^2 - |D|^2 + |N|^2$ & circular & along $z^\prime$   & -\\
$F \frac{d\sigma}{dt}$ &$2Re(S_1N^* + S_2D^*)$              & circular & along $x^\prime$   & -\\
\hline
\hline
\multicolumn{4}{c}{Beam and Recoil Baryon Polarization}\\
\hline
$O_{x'} \frac{d\sigma}{dt}$ &$-2Im(S_2D^* + S_1 N^*)$       & linear   & -                   & along $x^\prime$\\
$O_{z'} \frac{d\sigma}{dt}$ & $-2Im(S_2S_1^2 + N D^2)$      & linear   & -                   & along $z^\prime$\\
$C_{x'} \frac{d\sigma}{dt}$ & $-2Re(S_2N^* + S_1 D^*)$      & circular & -                   & along $x^\prime$\\
$C_{z'} \frac{d\sigma}{dt}$ & $|S_2|^2-|S_1|^2 -|N|^2+|D|^2$& circular & -                   & along $z^\prime$\\
\hline
\hline
\multicolumn{4}{c}{Target and Recoil Baryon Polarization}\\
\hline
$T_{x'} \frac{d\sigma}{dt}$ &$2Re(S_1S_2^* + N D^*)$         & -        & along $x^\prime$   & along $x^\prime$\\
$T_{z'} \frac{d\sigma}{dt}$ &$2Re(S_1N^* + S_2D^*)$          & -        & along $x^\prime$   & along $z^\prime$\\
$L_{x'} \frac{d\sigma}{dt}$ & $2Re(S_2N^* - S_1D^*)$         & -        & along $z^\prime$   & along $x^\prime$\\
$L_{z'} \frac{d\sigma}{dt}$ &$|S_1|^2+|S_2|^2-|N|^2-|D|^2$   & -        & along $z^\prime$   & along $z^\prime$\\
\hline
\end{tabular}
\caption{The set of all observables for pseudoscalar
  meson photoproduction using helicity amplitudes.  The meaning of the
  amplitudes $N$, $S_1$, $S_2$ and $D$ is discussed in the text.  
  The table is adapted from Ref.~\cite{bds}.}
\label{obstable}
\end{center}
\end{table*}

\subsection{$z$-Spin Basis Model}
\label{sec:zspin}

\begin{figure}
\includegraphics[scale=0.32,angle=-90.0]{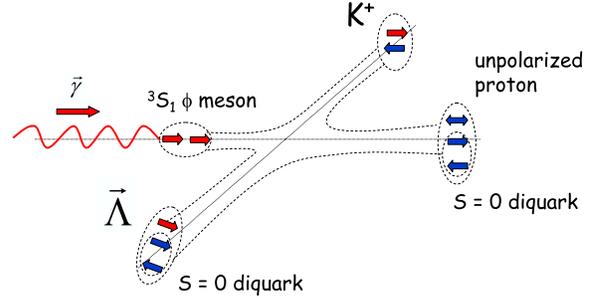}
\caption{ (color online) 
Quark-line picture of how full polarization
of the $\Lambda$ may be achieved.  An $\overline s s$ quark pair
produced from the photon hadronizes such that the $s$ quark in the
$\Lambda$ retains its full polarization after precessing due to
spin-orbit or spin-spin interactions, while the $\overline s$ quark
ends up in the spinless kaon.}
\label{fig:quarks}       
\end{figure}

In lieu of working with the four helicity amplitudes, we next consider
a simplified picture in which a spin 1/2 object scatters from a
spinless target~\cite{schumachercxcz}.  The ansatz for this picture is
that the incoming virtual photon fluctuates into a $^3S_1$ pair of
$\overline{s}s$ quarks, as shown in Fig.~\ref{fig:quarks}.  The
strange quark which determines the spin of the $\Lambda$ is formed
fully polarized, but during hadronization has its polarization
direction precessed.  This will be the case if the hadronization
process is of the spin-orbit or spin-spin form, because in both
quantum and classical physics such interactions preserve the magnitude
of a spin, but precess its direction.  In a $z$-spin basis, we can
model this with two amplitudes.  Let $g(\theta)$ represent the
amplitude for no flip of the $s$-quark spin as a function of
production angle $\theta$, and $h(\theta)$ be the amplitude that does
flip the spin.  The scattering matrix, $S$, that acts on an initial
two-component $s$-quark spinor $\chi_0$ in the $z$-spin basis, has the
form
\begin{equation}
S =
\left(
\begin{array}{cc}
g(\theta)           & h(\theta)e^{-i \xi} \\
-h(\theta)e^{i \xi} & g(\theta)
\end{array}
\right),
\label{eq:matrix}
\end{equation}
where $\xi$ is the azimuthal production angle with respect to a fixed
coordinate basis.  If the initial spin 1/2 density matrix is $\rho_0 =
\chi_0 \chi_0^\dagger$, then the polarization vector $\vec P_f$ of the
final state hyperon is evaluated using the spin operator $\vec\sigma$
via
\begin{equation}
\vec P_f = \frac{Tr(\rho_0 S^\dagger \vec\sigma S)}{Tr(\rho_0 S^\dagger  S)}
\label{eq:polfinal2}
\end{equation}
In terms of the amplitudes $g$ and $h$, the polarization vector
components can be written as
\begin{eqnarray}
|A|^2 =& g^*g + h^*h & 
\label{eq:dsdo} \\
P_{fx} =& \frac{g^*h + h^*g}{g^*g + h^*h} P_\odot &\equiv -C_x P_\odot 
\label{eq:pfx} \\
P_{fy} =& i \frac{g^*h - h^*g}{g^*g + h^*h}  &\equiv P 
\label{eq:pfy} \\
P_{fz} =& \frac{g^*g - h^*h}{g^*g + h^*h} P_\odot &\equiv C_z P_\odot
\label{eq:pfz}
\end{eqnarray}
where $P_\odot$ is the circular polarization of the incoming photon
beam.  Measurement of $\vec P_f$, plus knowledge of $P_\odot$ and the
cross section~\cite{bradforddsdo} then leads to a set of four
observables $P, C_x, C_z$, and $d\sigma/d\Omega$.

In order to make the amplitudes $g$ and $h$ dimensionless, we write
the cross section as a product of a phase space factor, dimensional
factors, and the magnitude of a dimensionless matrix element
$A(\theta)$.
\begin{equation}
\frac{d\sigma}{d\Omega} = \frac{1}{4}(\hbar c)^2 \alpha f
\frac{p_f}{p_i} \frac{1}{W^2}|A(\theta)|^2,
\end{equation}
where $\alpha$ is the fine structure constant, and $f$ is a
corresponding strong decay strength scale that was arbitrarily set
equal to 1 in this calculation.  $p_f$ and $p_i$ are the final and
initial state momenta in the reaction center of mass frame, and $W$ is
the invariant energy of the system.  The combined angle-independent
factors multiplying $A(\theta)$ range from 0.22 to 0.14 from low to
high values of $W$ in the results discussed below.  Using the four
experimentally determined quantities $|A|^2$, $C_x$, $C_z$, and $P$,
we can then solve for the magnitudes of $g$ and $h$, as well as the
phase difference $\Delta \phi$ between these two complex amplitudes.
The overall phase is unimportant.  The result is
\begin{eqnarray}
|g| &=& \left (\textstyle\frac{1}{2}|A|^2 (1+C_z)\right )^{1/2}, \\
|h| &=& \left (\textstyle\frac{1}{2}|A|^2 (1-C_z)\right )^{1/2}, \label{eq:aitch}\\
\Delta \phi &=& \tan^{-1} \frac{P}{C_x}  \\
   &=& \sin^{-1} \frac{P}{\sqrt{1-C_z^2}} =\cos^{-1} \frac{C_x}{\sqrt{1-C_z^2}} \label{eq:phase}.
\end{eqnarray}
The constraint of having four observables and three unknowns allows
for alternative ways of evaluating the phase difference $\Delta \phi$;
we pick the one for which the propagated measurement uncertainty is
most favorable.

\section{Results}

The CLAS measurements of $C_x$, $C_z$, and $P$ were combined to
construct the magnitude of the $\Lambda$ hyperon recoil polarization
when the hyperon is created from a circularly polarized photon beam.
Defining the magnitude as
\begin{equation}
|\vec R_\Lambda| \equiv \sqrt{P^2 + C_x^2 +C_z^2},  
\label{eq:are}
\end{equation}
it was expected that $|\vec R_\Lambda| < 1$ on the physical ground
that the photon spin polarization could be shared between the
recoiling hyperon and the orbital angular momentum contained in the
$K^+\Lambda$ final state.  If the final state were entirely $S$-wave,
as from the decay of an intermediate nucleon resonance such as the
$S_{11}(1650)$, then one would expect $|\vec R_\Lambda| = 1$.  But
there is no reason to expect the final state to be dominated by
$S$-wave, and indeed most hydrodynamics models include $P$- and
sometimes $D$- wave intermediate nucleon resonances.  It was
unexpected, therefore, when the overall average magnitude was found to
be
\begin{equation}
\overline{R}_\Lambda = 1.01 \pm 0.01,
\end{equation}
where the uncertainty is that of the weighted mean of the points in
Fig.~\ref{fig:lambda}.  The estimated systematic uncertainty is about
$\pm 0.03$.  A more complete presentation of the data is given in
Ref~\cite{bradford}.

The precision of the result is limited by the CLAS data on transferred
polarization $C_x$ and $C_z$.  The CLAS data for induced $P$ have
recently be seen to be in excellent agreement with results from
GRAAL~\cite{lleres}. But incorporating those values for $P$ into the
calculation of $\vec R_\Lambda$ does not lead to smaller
uncertainties.  Thus, a new experiment for $C_x$ and $C_z$ would be
needed to improve the overall precision of $\vec R_\Lambda$.

In electroproduction, the $K^+ \Lambda$ final state has been studied
in the range $0.3 < Q^2 < 1.5$ (GeV/c)$^2$, integrated over all kaon
production angles~\cite{carman}.  The electron beam was longitudinally
polarized, allowing extraction of the spin transfer to the hyperon in
a manner analogous to the photoproduction analysis.  The hyperon spin
was found to be dominantly in the direction of the photon momentum,
that is, along the $z$ axis of virtual photon.  The phenomenology away
from $Q^2 = 0$ was thus seen to be quite similar to the
photoproduction result emphasized in this paper.

The other recent measurements of polarization observables were for the
beam spin asymmetry, $\Sigma$, using linearly polarized photons.
Measurements from threshold up to 1.5 GeV photon energy made at
GRAAL~\cite{lleres} are in good agreement with measurements made at
LEPS/SPring8 ~\cite{zegers} at energies from 1.5 to 2.4 GeV.  The
observable $\Sigma$ can not be predicted on the basis of the
constraint implied by the CLAS results, unfortunately, as discussed
below.

\subsection{Helicity Amplitudes Constraint}
\label{sec:res_helicity}

Combining the experimental result discussed above with the helicity
amplitude relations summarized in Table~\ref{obstable}, we can derive
an expression relating the helicity amplitudes.  We find that
\begin{equation}
S_1 S_2 = N D.
\label{eq:ssnd}
\end{equation}
That is, the product of the two single-helicity-flip amplitudes is
equal to the product of the the no-flip and the double-flip
amplitudes.  This is the empirical constraint implied by the data.
The next step would be to make predictions for what this constraint
imposes on observables that have been (or soon will be) measured, such
as $\Sigma$, $O_x$, $O_z$, or $T$ ~\cite{lleres2,althoff}.  However,
this seems to be fruitless based on this constraint alone.  Note that
Eq.~\ref{eq:ssnd} is for products of amplitudes, while the observables
are constructed out of bilinear products of an amplitude and an
complex conjugate amplitude.  In an earlier
paper~\cite{schumachercxcz} we speculated that if a linearly polarized
beam were used to measure $O_x$ and $O_z$, then in combination with
$P$ one might again find a magnitude of the polarization vector to be
unity, that is,
\begin{equation}
O_x^2 + O_z^2 + P^2 = 1.
\label{eq:oop}
\end{equation}
But this does {\it not} follow; the constraint given by
Eq.~\ref{eq:ssnd} does not lead to this conclusion.  The conjecture
given by Eq.~\ref{eq:oop} may or may not be true, but the information
in hand does not predict either one outcome or the other.  There are
results coming from GRAAL for these observables~\cite{lleres2}, and so
we may have some insight in this area soon.

\subsection{$z$-Spin Basis Model}
\label{sec:res_zspin}

\begin{figure}[htbp]
\includegraphics[scale=0.35]{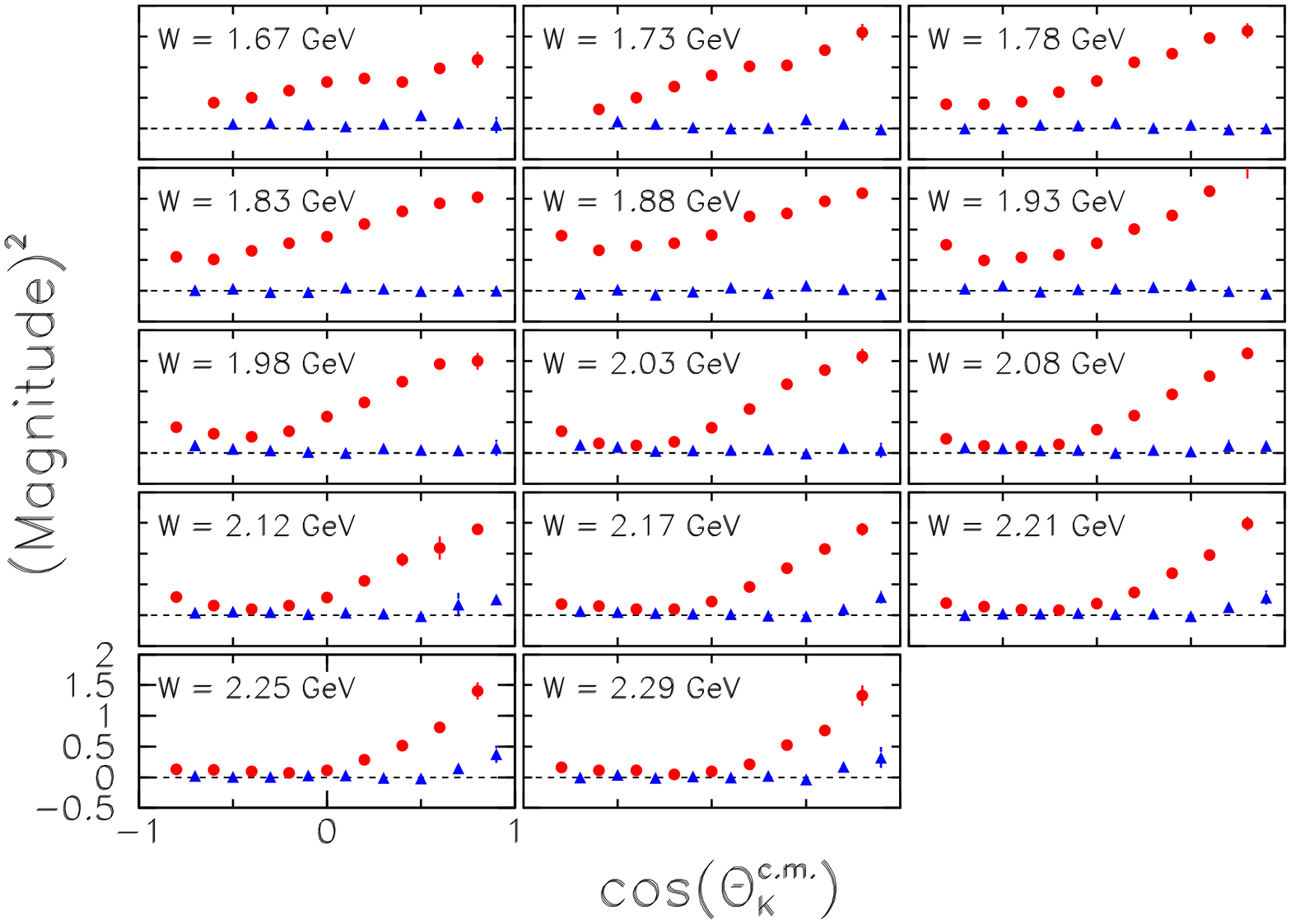}
\vspace{-1.0cm}
\caption{ 
(color online) The squared magnitudes of the spin non-flip
$|g|^2$ (red circles) and spin flip $|h|^2$ (blue triangles)
amplitudes for the reaction $\gamma + p \rightarrow K^+ + \Lambda$.
Each panel for a bin in $W$ is given as a function of c.m. kaon
production angle.  Note the displaced zero on the vertical axis, and
that the points are slightly shifted in angle to avoid overlaps.  
}
\label{fig:mags}       
\end{figure}

In terms of the two-amplitude model discussed in the previous section,
Fig.~\ref{fig:mags} shows the squared-magnitudes of the amplitudes
$g(\theta)$ and $h(\theta)$. The squares are shown since, as seen in
Eq.~\ref{eq:aitch}, $|h|^2$ can be negative in regions where the $C_z$
measurement exceeds the physical limit of $C_z=1$ due to fluctuations.
This representation amplifies the main message: the non-flip amplitude
$g$ is the dominant one at forward angles, while the spin-flip
amplitude $h$ plays a significant role mainly at higher energies in
the backward hemisphere.  This reaction is occurring very far from the
limit of quark helicity conservation, and is indeed much closer to the
limit of $s$-channel spin conservation at the baryon level.

\begin{figure}[htbp]
\includegraphics[scale=0.35]{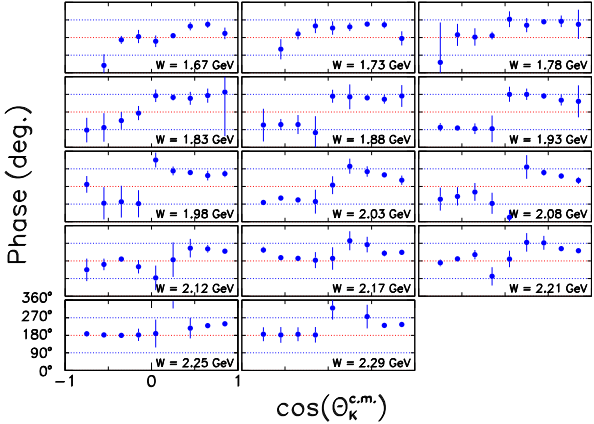}
\vspace{-1.0cm}
\caption{ 
(color online) 
The phase difference of the spin non-flip $g$
and spin flip $h$ amplitudes, $\Delta \phi = \phi_g - \phi_h$, for the
reaction $\gamma + p \rightarrow K^+ + \Lambda$.  The binning and the
various $W$ panels match those of the previous figure.  
}
\label{fig:phase}       
\end{figure}

The relative phase, $\Delta \phi = \phi_g - \phi_h$ is shown in
Fig.~\ref{fig:phase}.  For each datum, the form of Eq.~\ref{eq:phase}
that resulted in the smallest propagated error was adopted.
Qualitatively, we see that with increasing $W$ the relative phase
evolves from being nearer $\pm \pi/2$ to being nearer to $\pi$.  The
discontinuities seen near $W=2.00$ GeV arise when the value of $P$
passes through zero while $C_x$ remains small and negative.  The
trends are of limited statistical significance, unfortunately.  There
are no model-based expectations available with which to compare the
results of this simple two-amplitude model.  It's main value is to
underscore the dominance of the spin non-flip nature of this reaction.

\subsection{Semi-Classical Model}
\label{sec:classical}

In a magnetic field, the expectation value of a quantum mechanical
spin with magnetic moment $\vec \mu$ evolves in time the same way as a
classical ``spin'' angular momentum vector does~\cite{merz}.  In a
constant magnetic field, a spin $\vec\mu$ will precess at its Larmor
frequency due to a torque $\vec \mu \times \vec B$.  Starting from
this observation, one can build a qualitative picture of the reaction
mechanism discussed in this paper.  We take literally the head-to-tail
triplet orientation of the initial state configuration of a virtual
$\overline{s} s$ quark pair shown in Fig.~\ref{fig:quarks}.  Their
spin-spin dipole interaction is taken to be of the classical form
$\vec \mu\cdot\vec B$, where the field of one dipole at the location
of the other at relative location $\vec r$ is $\vec
B=\frac{1}{r^3}\left[ 3(\vec\mu\cdot\hat r)\hat r - \vec \mu \right]$.
The strength scale of the field is not that of the electromagnetic
interaction, since we expect the color-magnetic (gluonic) force to be
dominant.  Indeed, in the result shown below the value of the
effective fine structure constant $\alpha_{eff}$ was increased by a
factor of 30 to get qualitatively reasonable behavior.  The dipole
strength of the strange quark was written as $|\vec\mu| = \alpha_{eff}
\frac{(\frac{2}{3}e)\hbar}{2m_s}\sqrt{\frac{1}{2}(\frac{1}{2}+1)}$
where the strange quark mass, $m_s$, was taken as 150 MeV.  The
spacing of the initial-state quarks was set to half the wavelength of
the incoming photon in the overall c.m. frame.

As long as the quark pair remains axially aligned there is no
dipole-dipole torque.  But then one considers the motional (color-)
magnetic field of a proton charge distribution as it interacts with
the quark pair.  Modeling a proton as a spinless charge distribution
with a root-mean-square radius of $R_{rms}=0.86$ fm, the collision of
moving proton and quark magnetic dipoles first precesses the dipoles
off their alignment axis.  The spin-spin interaction then serves to
precess the quarks further.  The semi-classical model system can be
allowed to interact in this way for some characteristic
``hadronization time'' which was taken to be 1 fm/$v_{cm}$, where
$v_{cm}$ is the speed of the quark pair in the overall c.m. system.
After this ``hadronization time'' the strange quark is presumed to
have formed a $\Lambda$ hyperon, to have moved out of range of the
strong force, hence freezing its orientation.  The production-angle
dependence of the model is built into a mapping of the classical
collisional impact parameter, $b$, onto scattering angle.  Parameter
$b$ affects the motional (color-) magnetic field and hence degree of
quark precession.  We used a Rutherford-like mapping $\theta = 2
\tan^{-1}(2b/r_0)$ where $r_0 = R_{rms}/\sqrt{12}$.

\begin{figure}
\includegraphics[scale=0.30]{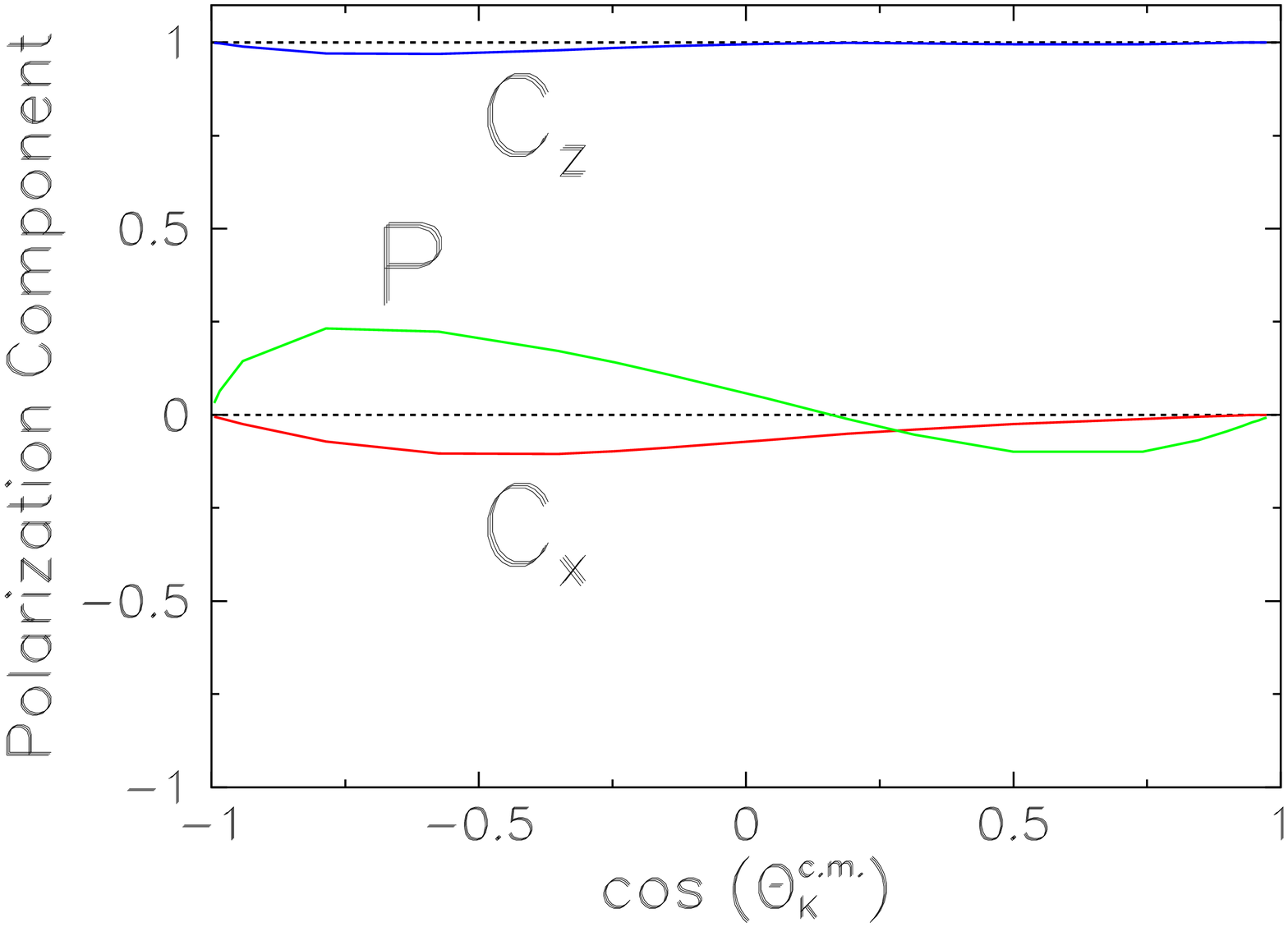}
\vspace{0.0cm}
\caption{ 
(color online) 
Results of a semi-classical model of $\Lambda$ hyperon polarization for a
total c.m. energy of  $W=2$
GeV.  The three projections of the polarization vector are shown as a
function of kaon production angle.  The model is based on the
spin-spin and spin-orbit interaction of an $\overline s s$ quark pair
and the (color-) magnetic field of a proton charge distribution.
}
\label{fig:quarktorque}       
\end{figure}

A numerical simulation was developed to compute the resultant hyperon
polarization as a function is scattering angle.  The result is shown
in Fig.~\ref{fig:quarktorque} for the typical case of $W=2.0 $ GeV. It
was remarkably simple to find a set of model parameters that mimic the
observed phenomenology.  That is, $C_z$ is large and positive, meaning
that the strange quark spin was not much perturbed from its initial
direction.  $C_x$ remains small and negative across the range of
production angles.  Finally, the out-of-plane polarization component
$P$ is negative at forward production angles and positive and backward
production angles.  This is what is seen in the
measurements~\cite{bradford}.  By construction, $\vec R_\Lambda = 1$
in this model.

We cannot expect this semi-classical picture to be developed into a
realistic model of this reaction.  There are many free parameters in
this model that cannot be connected in any useful way to measured
properties of the particles involved.  Its main value is to offer a
possible physical explanation of how, classically, a spin can be
created and made to partially point ``out of plane'' with respect to
the beam polarization axis.  It offers heuristic support to the
hypothesis that the observed $\Lambda$ polarization arises from the
creation of a strange quark pair in a triplet state that is acted upon
by a spin-magnitude preserving color-magnetic hadronization process.

\section{Further Discussion and Conclusions}
\label{sec:3}

The CLAS results on $C_x$, $C_z$, and $P$ have recently been
satis\-factor\-ily fit in coupled-chan\-nel baryon-reso\-nance mo\-del by
the Bonn-Gachina group~\cite{bg}.  In that model, the quark-level
dynamics advocated here are not relevant, but instead a
well-defined set of $s$-channel isobars are made to interfere to
explain the $K^+\Lambda$ results, together with a number of other
reaction channels.  A new $P_{13}$ resonance near 1860 MeV was
introduced to fit the $C_x$ and $C_z$ results.  Their approach could
be the most profitable one for explaining the broad array of reaction
channels in terms of a unified model framework.  But it is worth
keeping in mind that the apparently ``simple'' result that $|\vec
R_\Lambda| = 1$ across energy and angle may signal a more fundamental
dynamic about how the reaction mechanism proceeds, something other
than the interference of multiple baryon resonances.

In this discussion we have ignored the less-precise data available for
the polarization components of the $\Sigma^0$
hyperon~\cite{bradford,mcnabb}.  In the flavor SU(3) quark model, the
magnetic moment of the $\Sigma^0$ is opposite to that of the
$\Lambda$, and indeed it is seen that observables $P$ for the two
hyperons are roughly opposites of each other.  The $C_z$ and $C_z$
behaviors are not similar, but the data are consistent with the
requirement that $C_z$ for the $\Sigma^0$ goes to unity at forward
angles.  The phenomenological models discussed in this paper have yet
to be adapted to the $\Sigma^0 $ case.

We have discussed the observables available in hyperon
photoproduction, in particular $K^+ \Lambda$ photoproduction from the
proton, in light of recent measurements by the CLAS collaboration of
$C_x$, $C_z$, and $P$.  In the helicity basis, one obtains a product
relationship among the four amplitudes.  By itself, this empirical
constraint does not lead algebraically to predictions for the other
spin observables.  But the observation that the $\Lambda$ hyperon is
produced fully polarized from a photon of definite helicity suggests a
physical picture in which a virtual $s$ quark is created with full
polarization, and that the hadronization process preserves the full
magnitude but not the direction.  In a two-amplitude model in a
$z$-spin basis, the result says that the spin non-flip amplitude is by
far the dominant one over the spin-flip amplitude.  We computed the
magnitudes and relative phase of these amplitudes, which showed smooth
trends across energy and production angle.  A semi-classical model of
spin-spin and spin-orbit interactions of (color-) magnetic moments
with their associated fields was seen to reproduce the qualitative
features of the data.  Further work to interpret these results is
needed, as well as additional constraints from other observables.

In closing, it is encouraging to note that experimental work at
CLAS~\cite{frost}, Crystal Barrel/TAPS at Bonn~\cite{bonn}, and
GRAAL~\cite{lleres2} is underway to measure additional spin
observables in hyperon photoproduction and related reactions in
pseudoscalar meson photoproduction.  Therefore, progress in unraveling
the amplitude-level structure of these reactions can be expected.  It
will be interesting to see to what extent the physical mechanisms
discussed in this paper will be supported.

%

\end{document}